\documentclass[10pt,twocolumn]{article}

\usepackage[T1]{fontenc}
\usepackage[utf8]{inputenc}
\usepackage[margin=0.75in,columnsep=0.25in]{geometry}
\usepackage{graphicx}
\usepackage{amsmath,amssymb}
\usepackage{amsthm}
\usepackage{booktabs}
\usepackage{array}
\usepackage{authblk}
\usepackage{tikz}
\usetikzlibrary{arrows.meta,positioning}
\usepackage{xurl}
\usepackage[hidelinks]{hyperref}
\usepackage{orcidlink}
\usepackage[backend=biber,style=ieee,sorting=none]{biblatex}
\newcommand{\code}[1]{\texttt{#1}}
\begin{document}

\title{A Controlled Candidate-Set Benchmark for Offline Satellite-Security Plan Decomposition}

\author[1,2]{João Paolo Cavalcante Martins Oliveira\,\orcidlink{0000-0003-4117-953X}}
\author[3]{Lucas Teske\,\orcidlink{0009-0002-8526-7662}}
\author[4]{Paulo Matias\,\orcidlink{0000-0002-6504-5141}}
\affil[1]{Universidade Federal do Rio Grande do Norte -- UFRN\\
  \texttt{paolo.oliveira.948@ufrn.edu.br}}
\affil[2]{SETI Institute}
\affil[3]{TeskesLab\\\texttt{lucas@teske.com.br}}
\affil[4]{Universidade Federal de São Carlos\\\texttt{matias@ufscar.br}}
\date{}

\maketitle

\begin{abstract}
Some security-development settings require local models that map an objective to an
ordered, checkable plan. We present a low-rank \emph{decomposition adapter} and release
a case-disjoint corpus with 24 authored decompositions and 83 derived next-step examples
across 24 satellite-security cases. To prevent reference-plan leakage, prompts contain one
objective and eight shuffled SPARTA
candidates. The evaluation candidate sets are constructed with oracle inclusion of every
reference technique, so this study measures selection from a controlled set, not retrieval.
On six fixed cases and five decoding seeds, mean scores vary across adapters and prompted
Qwen2.5 baselines; a separate greedy check over three adapter training seeds shows
non-negligible training variance. A 24-fold greedy leave-one-case-out control at the two
compact sizes likewise shows tradeoffs: at 1.5B, adapter precision is 0.583 against 0.480
for two-shot prompting, while recall is 0.586 against 0.660. Case-resampling intervals for adapter
precision relative to the strongest prompted baseline cross zero at 1.5B and 7B. The
adapter is also substantially more format-compliant, which prevents attributing all score
differences to selection. Autoregressive next-step accuracy is 0.08, 0.06, and 0.29 at
0.5B, 1.5B, and 7B. We therefore offer the formulation, leakage-controlled dataset,
reproducibility package, and descriptive analysis as a proof of concept, not evidence of a
reliable stand-alone decomposition system or a general adapter advantage. A separate
diagnostic tests contract reuse.
\end{abstract}


\section{Introduction}
\label{sec:intro}

A development-time satellite-security testbed may need to map an objective, such as
testing telecommand authorization, to ordered steps tied to recognized techniques and
checks. When sensitive artifacts cannot leave a local environment, a compact local model
is an attractive design option. Decomposition prompting can improve compositional
reasoning~\cite{leasttomost2023}, and task specialization can transfer multi-step behavior
to smaller models~\cite{smallreasoning2023}; neither result establishes that the same
strategy works for security-campaign plans.

We study a narrower component: a low-rank adapter~\cite{lora2022} trained to emit a plan
conditioned on an explicit candidate set. Such a set could eventually come from a
retriever, but the present evaluation constructs it with oracle inclusion of the reference
techniques. The supervised completions contain concrete techniques and actions, and their
loss is conditioned on the prompt, so this design neither excludes memorization nor
demonstrates factual updating.

This paper isolates that front end from a larger grounded, verifiable red-teaming
testbed so the decomposition component can be formalized, released, and measured on its
own. We provide:

\begin{itemize}
  \item a formulation of candidate-set objective decomposition in which an auditable
        technique set conditions a completion-masked low-rank adapter (Section~\ref{sec:form},
        Section~\ref{sec:method});
  \item the released SatSec corpus, with 24 authored decompositions and 83 derived
        next-step examples across 24 documented and canonical cases, split disjointly
        by case, with two supervision modes (full decomposition and next-step) and an
        authored check intended to confirm every step~\cite{satsecdataset2026}
        (Section~\ref{sec:method});
  \item a descriptive evaluation on controlled candidate sets that mix reference
        techniques with cross-family distractors, using
        completeness (recall), precision (selectivity), ordering, candidate validity, and
        check-presence metrics, paired case-level intervals, ordering-support counts,
        and formatting diagnostics at 0.5B, 1.5B, and 7B
        and a separately bounded reverse-engineering transfer diagnostic
        (Section~\ref{sec:eval}).
\end{itemize}

This is a defensive, development-time artifact. The corpus contains concrete technique
identifiers and actions and is intended only for authorized or emulated
targets in an offline setting under responsible disclosure, the same scope as the
testbed it feeds.

\section{Problem Formulation}
\label{sec:form}

Let an objective $o$ be the root of an attack tree. Its reference decomposition is an
ordered sequence
\begin{equation}
  \sigma(o) = (s_1, s_2, \dots, s_{n_o}),
\end{equation}
where each sub-step is a triple
\begin{equation}
  s_i = (t_i,\ \tau_i,\ c_i),
\end{equation}
with $t_i$ a natural-language action, $\tau_i \in \Sigma$ a technique identifier from
the SPARTA technique set $\Sigma$~\cite{sparta}, and $c_i$ an authored check statement
intended to confirm $s_i$. Whether $c_i$ is actually deterministic and sufficient is a
semantic-review question, not an automatic metric. A candidate-set function $G(o) \subseteq \Sigma$
provides the techniques available to the model. In this study, $G(o)$ is constructed to
contain every reference technique in $\sigma(o)$ together with in-domain distractors; it
is therefore an oracle candidate set, not the output of an evaluated retriever. A useful policy must
\emph{select} the reference techniques out of $G(o)$ and \emph{order} them, not emit the
whole candidate set. Precision (selectivity) is what distinguishes selection from dumping,
and we measure it alongside recall in Section~\ref{sec:eval}.

The learning target is a policy $p_\theta$ that maps a grounded objective to a
decomposition. We supervise two modes. The \emph{full decomposition} mode factorizes
autoregressively,
\begin{equation}
  p_\theta\big(\sigma \mid o, G(o)\big)
     = \prod_{i=1}^{n_o} p_\theta\big(s_i \mid o, G(o), s_{<i}\big),
  \label{eq:decompose}
\end{equation}
and the \emph{next-step} mode supervises a single continuation,
\begin{equation}
  p_\theta\big(s_k \mid o, G(o), s_{<k}\big).
  \label{eq:nextstep}
\end{equation}
The next-step mode densifies supervision along the campaign and targets the
long-horizon coherence failure directly, by turning one campaign into many local
prediction problems.

The design principle is to keep the candidate set explicit and inspectable while the model
emits a structured selection. Whether a practical retriever can supply a useful $G(o)$ is
outside the present experiment.

The case-disjoint evaluation reduces direct case reuse, while the limitations of a small
corpus and shared technique vocabulary remain explicit (Section~\ref{sec:threats}).

\section{The Decomposition Adapter}
\label{sec:method}

\begin{figure*}[t]
\centering
\resizebox{0.96\textwidth}{!}{%
\begin{tikzpicture}[
  x=1cm,
  y=1cm,
  font=\footnotesize,
  box/.style={draw, rounded corners, align=center, minimum height=9mm,
              text width=2.45cm, inner sep=3pt},
  data/.style={box, fill=black!4},
  process/.style={box, fill=black!9},
  artifact/.style={box, fill=black!15},
  future/.style={box, dashed, fill=white},
  arrow/.style={-{Latex[length=2mm]}, thick},
  futurearrow/.style={-{Latex[length=2mm]}, dashed}]

  \node[anchor=east, font=\bfseries] at (-0.35,1.1) {Training};
  \node[data]     (corpus)   at (1.2,1.1) {24 authored cases\\and SPARTA records};
  \node[process]  (builder)  at (4.4,1.1) {Deterministic builder\\and leakage audit};
  \node[data]     (examples) at (7.6,1.1) {Case-disjoint dataset\\82 train, 25 test};
  \node[process]  (loss)     at (10.8,1.1) {Completion-only LoRA\\train split only};
  \node[artifact] (adapter)  at (14.0,1.1) {Trained low-rank\\adapter};

  \draw[arrow] (corpus) -- (builder);
  \draw[arrow] (builder) -- (examples);
  \draw[arrow] (examples) -- (loss);
  \draw[arrow] (loss) -- (adapter);

  \node[anchor=east, font=\bfseries, align=right] at (-0.35,-1.7)
        {Controlled\\evaluation};
  \node[data]     (objective) at (1.2,-1.7) {Held-out objective $o$};
  \node[data, text width=2.8cm] (candidates) at (4.6,-1.7)
        {Oracle candidate set $G(o)$\\eight shuffled techniques};
  \node[process, text width=2.8cm] (model) at (8.2,-1.7)
        {Frozen Qwen2.5\\$+$ trained adapter};
  \node[data]     (plan) at (11.6,-1.7)
        {Structured plan\\action, technique, check};
  \node[process]  (score) at (14.8,-1.7)
        {Offline scoring\\R, P, O, G, C};

  \draw[arrow] (objective.north) -- ++(0,0.5) -| (model.north west);
  \draw[arrow] (candidates) -- (model);
  \draw[arrow] (adapter.south) -- (model.north east);
  \draw[arrow] (model) -- (plan);
  \draw[arrow] (plan) -- (score);

  \node[future, text width=2.8cm] (retriever) at (4.6,-3.9)
        {Future retriever\\not evaluated};
  \node[future] (verify) at (11.6,-3.9)
        {Future verification\\not evaluated};
  \draw[futurearrow] (retriever) -- (candidates);
  \draw[futurearrow] (plan) -- (verify);
\end{tikzpicture}
}
\caption{Architecture and experimental boundary. Solid boxes and arrows are implemented in
the released training and evaluation pipeline. Evaluation uses controlled oracle candidate
sets containing every reference technique, so it measures selection and ordering rather than
retrieval. Dashed boxes show possible deployment components; neither retrieval nor semantic
verification is evaluated here. R, P, O, G, and C denote recall, precision, ordering,
candidate validity, and check-field presence.}
\label{fig:architecture}
\end{figure*}
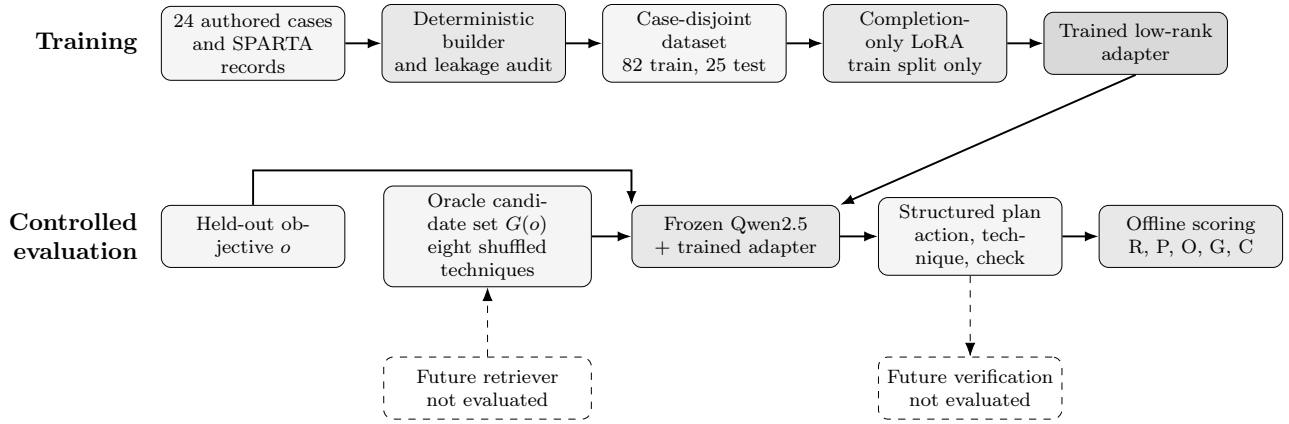

Figure~\ref{fig:architecture} separates the adapter-training path from the controlled
evaluation path and from possible deployment components that are outside this study.

\subsection{Data construction}
\label{sec:data}

The corpus is built deterministically from an attack-tree library grounded in SPARTA
and space-security standards. Each case supplies an objective $o$, candidates $G(o)$,
a reference $\sigma(o)$, and a check per step. We emit two chat-formatted example types.
For \code{decompose}, the prompt contains $(o,G(o))$ and the completion is $\sigma(o)$
(Eq.~\eqref{eq:decompose}). For \code{next\_step}, the prompt also contains $s_{<k}$
and the completion is $s_k$ (Eq.~\eqref{eq:nextstep}).

The released set holds $|\mathcal{D}| = 107$ examples over $24$ cases, comprising
$24$ author-written \code{decompose} examples and $83$ mechanically derived
\code{next\_step} prefixes. Cases span documented incidents and controlled research,
including satellite-link abuse and passive eavesdropping~\cite{turla2015,pavur2020whispers},
GNSS spoofing~\cite{bhatti2017hostile}, and unauthenticated telecommand
analysis~\cite{willbold2023space}, plus canonical standards-grounded and
development-bench patterns~\cite{ccsds355,sparta}.
Because SPARTA is a living framework, the artifact does not invent a release version:
its local page titles and identifiers were audited against the live site on 2026-07-27,
with the corrected GNSS mapping rechecked on 2026-07-26.
Splitting is \emph{disjoint by case}: the cases are partitioned so no case's steps
appear in both train and test, with the evaluation cases held out entirely. The released
snapshot fixes a split with six cases held out for evaluation, giving $82$ train and $25$
test examples.

To prevent input leakage, the reference narrative is removed from every input. The objective field is one
high-level sentence and contains no step sequence; each candidate block contains eight
shuffled candidates. An automated audit rejects legacy sequence markers, future-step
titles in next-step prompts, split overlap, malformed grounding, and gold plans that do
not self-score perfectly. We also corrected the GNSS mapping: EX-0002 is PNT geofencing,
whereas the counterfeit-signal and controlled carry-off step maps to EX-0014.04.

To make the task a selection problem rather than a copy, each candidate block $G(o)$ is
padded to eight distinct techniques: the reference techniques plus plausible in-domain
distractors drawn deterministically (a per-case seed) from the SPARTA technique pool. Any
technique that shares a parent with a reference is excluded, so a distractor is a distinct
in-domain technique from another family, not a sub-technique sibling of a reference. This
defines the present benchmark as cross-family selection; it does not test sibling-level
near-miss discrimination (Section~\ref{sec:threats}). The combined block is then
shuffled, so candidate order carries no signal about which techniques are the reference ones
or in what order they belong. The reference plan is unchanged: it names only the reference techniques, in order, so
the model is supervised to pick them out of the noisy block and sequence them. A non-learned
baseline that emits every identifier in $G(o)$ therefore scores full recall but low precision
(0.40) and low ordering (0.44), which fixes the selectivity and ordering bar the adapter must
clear (Section~\ref{sec:eval}). Reference plans and checks were assembled and collectively
reviewed by the author team against the cited case sources and SPARTA mappings. They are
benchmark operationalizations, not independently certified ground truth; all automatic
scores below measure agreement with them. The artifact makes that construct inspectable:
all 24 cases resolve to nonempty source records, every reference step has Action and Check
fields, every reference technique occurs in its supplied candidate set, and all 34 unique
reference identifiers passed a live SPARTA existence/title audit. Stable evidence-scope notes
separate source-supported mechanisms from author-created boundaries, mappings, orders, and
checks.

\subsection{Low-rank, completion-only supervision}
\label{sec:train}

We install the decomposition behavior with a low-rank adapter~\cite{lora2022}. For a
frozen pretrained weight $W_0 \in \mathbb{R}^{d\times k}$, the adapted weight is
\begin{equation}
  W = W_0 + \frac{\alpha}{r}\, B A,
  \qquad B \in \mathbb{R}^{d\times r},\ A \in \mathbb{R}^{r\times k},
  \label{eq:lora}
\end{equation}
with rank $r \ll \min(d,k)$ and scaling $\alpha$. Only $\{A,B\}$ are trained; $W_0$
stays frozen, producing a smaller trained artifact than full-model finetuning.

Training uses a \emph{completion-only} objective. An example is a prompt token sequence
$x_{1:p}$ (system and user, including the candidate set) and a completion
$y_{1:m}$ (the assistant turn). The loss masks the prompt,
\begin{equation}
  \mathcal{L}(\theta) =
    - \mathbb{E}_{(x,y)\sim\mathcal{D}}
      \sum_{j=1}^{m} \log p_\theta\big(y_j \mid x, y_{<j}\big),
  \label{eq:loss}
\end{equation}
so prompt token positions are excluded from the token-level loss. The grounding $G(o)$
still determines the hidden states used to predict $y$, and loss gradients backpropagate
through those context-conditioned computations into the adapter. Moreover, supervised
completions contain concrete identifiers, actions, and checks. Completion masking is
therefore a training choice, not a facts-out-of-weights or no-memorization guarantee.

\subsection{Grounded inference}
\label{sec:infer}

In a future deployment, retrieval could supply $G(o)$ for the objective's surface; the
present experiment instead uses the controlled oracle set described above. In
\code{decompose} mode the adapter emits the full ordered plan; in
\code{next\_step} mode it proposes the next step given the steps so far, which lets an
orchestrator interleave planning with execution and re-grounding. The requested schema asks
each emitted step to name a technique drawn from $G(o)$ and a deterministic check. The model
does not always comply, so identifier validity and check-field presence are measured rather
than assumed. A parsed plan can be passed to a separate verification layer, which we do not
evaluate. We also do not test whether
the adapter generalizes to newly added techniques without retraining.

\section{Evaluation Protocol}
\label{sec:eval}

We evaluate a fixed split that holds out six cases entirely. The corpus is small, so the
numbers characterize these unseen cases rather than estimate a population.

\subsection{Setup}
The base models are Qwen2.5-0.5B-Instruct, Qwen2.5-1.5B-Instruct, and
Qwen2.5-7B-Instruct at immutable Hugging Face revisions
\code{7ae5576}, \code{989aa79}, and \code{a09a354}, respectively. Each adapter uses
rank 16, alpha 32, dropout 0.05, no bias, and the \code{q}, \code{k}, \code{v},
\code{o}, \code{gate}, \code{up}, and \code{down} projection modules. Training uses
10 epochs, per-device batch size 1, gradient accumulation 8, learning rate
$2\!\times\!10^{-4}$, AdamW, a linear schedule without warmup, maximum sequence length
2048, bfloat16, and no quantization. The primary stochastic comparison uses the seed-42
adapter; a separate stability control trains seeds 42, 43, and 44 and evaluates each
greedily. Training ran in a container on one laboratory GPU per run. We study 0.5B and
1.5B as compact sizes and 7B as a scale comparison, but report no
on-device memory, latency, or throughput measurement. The
split is disjoint by case (Section~\ref{sec:data}). The six fixed cases are chosen to
span distinct attack families rather than a single behavior: passive eavesdropping (Pavur),
covert downlink abuse (Turla), unauthenticated telecommand (Space Odyssey), debug-interface
firmware extraction (JTAG), GNSS spoofing, and telecommand replay without link
authentication. This leaves 82 training and 25 test examples. These cases were inspected
during leakage and mapping corrections, so they are development-facing fixed cases, not
an untouched final test set. All configurations see the same oracle candidate set
(Section~\ref{sec:data}). We compare four inference
configurations at identical candidates: the unadapted instruction model with candidates only, with the
output schema spelled out in the prompt, the base with two in-context decomposition
exemplars drawn from training cases (never the case under test), and the base with the
decomposition adapter. The two exemplar cases are selected once with exemplar seed 0 and
held fixed across decoding seeds; an initial condition that changed exemplars together with
the decoding seed is retained only as a confounding diagnostic in the artifact.
Generation uses top-$p$ sampling with temperature 0.7 and $p=0.95$, repeated over five
decoding seeds. Table~\ref{tab:protocol} reports mean$\,\pm\,$standard deviation across
those repeated generations; this is sampling variation, not uncertainty across cases or
training runs. A separate exploratory analysis first averages generations within each case
and then resamples the six paired cases. Scoring is deterministic and separate from generation.
The GitHub package records the full prompts, selected few-shot cases, raw outputs,
generation settings, dataset hash, software versions, run manifests, and immutable
links to the separately hosted adapter weights.

\subsection{Metrics}
For a predicted decomposition against the authored reference $\sigma(o)$ we measure
reference agreement:
\begin{itemize}
  \item \textbf{Recall}: fraction of reference sub-steps recovered, a
        sub-step counting as recovered when the plan emits a step naming the same SPARTA
        technique. The paraphrastic natural-language check is not part of this automatic
        match.
  \item \textbf{Precision (selectivity)}: fraction of the distinct techniques the plan
        emits that are in the reference set. It penalizes copying distractors out of
        $G(o)$ and is the complement to completeness under the selection task: a policy that
        dumps the whole candidate set maxes recall but not precision.
  \item \textbf{Ordering fidelity}: one minus the rate of precedence violations against
        the reference order, computed over the steps whose technique matches the reference.
        Because it is defined on matched steps, values based on low-recall plans require
        caution.
  \item \textbf{Candidate validity}: fraction of emitted identifiers that are both valid
        SPARTA entries and drawn exactly from $G(o)$. Parent and sub-technique identifiers
        are distinct; a sibling never receives credit through a shared parent.
  \item \textbf{Check-field presence}: fraction of parsed steps carrying a non-empty
        \code{Check} field. This syntactic metric does not assess executability, safety,
        determinism, or whether the check confirms the action.
\end{itemize}

\subsection{Comparisons and results}
\textbf{Q1 (candidate selection) and Q2 (ordered decomposition).}
Table~\ref{tab:protocol} gives descriptive means on the leakage-controlled fixed cases. At
0.5B, adapter precision is 0.27 against 0.20 for fixed-exemplar two-shot, while recall is
0.26 against 0.28 and candidate-set validity is higher. At 1.5B, adapter precision is 0.44 against 0.35 for
schema prompting, but recall is 0.43 against 0.57. At 7B, adapter precision is 0.67
against schema's 0.59, while ordering is 0.68 against 0.84. These means are not evidence
of a general adapter advantage.

Paired six-case analysis reinforces that caution. Adapter-minus-schema precision has
a 95\% case-resampling interval of $[-0.034,0.211]$ at 1.5B and
$[-0.143,0.248]$ at 7B. Relative to two-shot prompting, the 0.5B candidate-validity interval is
$[0.259,0.506]$ and the 7B precision interval is $[0.222,0.305]$, but these comparisons
do not dominate prompted configurations across metrics. The adapter is also more
format-compliant: at 1.5B, 96.7\% of adapter predictions contain complete structured
fields, against 30.0\% for schema and 36.7\% for fixed-exemplar two-shot prompting. A format-neutral
diagnostic extracts the first occurrence of every SPARTA identifier anywhere in an output.
It raises the 1.5B schema precision from 0.35 to 0.36 and adapter precision from 0.44 to
0.44, but produces larger changes elsewhere: at 7B, candidate-only recall/precision rise
from 0.13/0.09 to 0.74/0.47. Thus structured parsing substantially understates some base
outputs, while the adapter's 1.5B and 7B precision remains 0.44 and 0.67 under the
format-neutral view. Format learning is a real component of the observed differences and
must not be conflated with semantic plan quality.

A reference-dependence diagnostic omits each fixed case in turn and recomputes paired means.
Adapter-minus-schema precision remains positive after every single-case omission, ranging
from $[0.042,0.185]$, $[0.045,0.130]$, and $[0.046,0.185]$ at 0.5B, 1.5B, and 7B,
respectively. Other metric/baseline signs are not uniformly stable. This shows that the
precision direction is not caused by one authored fixed-case reference; it does not turn the
six cases into a population sample or establish an overall adapter advantage.

\begin{table}[t]
\centering
\caption{Decomposition results over six development-facing fixed cases and five decoding seeds
(mean$\,\pm\,$sample std across decoding seeds). R/P/O/G/C are recall, precision,
ordering, candidate-set validity, and check-field presence. Two-shot exemplars are fixed.}
\label{tab:protocol}
\renewcommand{\arraystretch}{0.95}
\setlength{\tabcolsep}{2pt}
\scriptsize
\begin{tabular}{lccccc}
\toprule
\textbf{Config} & \textbf{R} & \textbf{P} & \textbf{O} & \textbf{G} & \textbf{C} \\
\midrule
\multicolumn{6}{l}{\emph{Qwen2.5-0.5B-Instruct}}\\
candidates & $.01\!\pm\!.02$ & $.02\!\pm\!.04$ & $.03\!\pm\!.07$ & $.03\!\pm\!.07$ & $.01\!\pm\!.01$ \\
schema    & $.19\!\pm\!.13$ & $.15\!\pm\!.11$ & $.36\!\pm\!.28$ & $.33\!\pm\!.16$ & $.92\!\pm\!.10$ \\
two-shot (fixed) & $.28\!\pm\!.08$ & $.20\!\pm\!.05$ & $.50\!\pm\!.12$ & $.59\!\pm\!.06$ & $.93\!\pm\!.05$ \\
$+$adapter & $.26\!\pm\!.07$ & $.27\!\pm\!.05$ & $.73\!\pm\!.15$ & $.98\!\pm\!.02$ & $1.00\!\pm\!.00$ \\
\midrule
\multicolumn{6}{l}{\emph{Qwen2.5-1.5B-Instruct}}\\
candidates & $.02\!\pm\!.04$ & $.01\!\pm\!.02$ & $.07\!\pm\!.15$ & $.01\!\pm\!.03$ & $.00\!\pm\!.01$ \\
schema    & $.57\!\pm\!.15$ & $.35\!\pm\!.09$ & $.71\!\pm\!.13$ & $.82\!\pm\!.11$ & $.93\!\pm\!.04$ \\
two-shot (fixed) & $.32\!\pm\!.24$ & $.22\!\pm\!.16$ & $.38\!\pm\!.22$ & $.66\!\pm\!.32$ & $.69\!\pm\!.34$ \\
$+$adapter & $.43\!\pm\!.04$ & $.44\!\pm\!.04$ & $.71\!\pm\!.10$ & $.91\!\pm\!.06$ & $.98\!\pm\!.04$ \\
\midrule
\multicolumn{6}{l}{\emph{Qwen2.5-7B-Instruct}}\\
candidates & $.13\!\pm\!.09$ & $.09\!\pm\!.08$ & $.17\!\pm\!.12$ & $.23\!\pm\!.18$ & $.11\!\pm\!.12$ \\
schema    & $.70\!\pm\!.06$ & $.59\!\pm\!.07$ & $.84\!\pm\!.08$ & $1.00\!\pm\!.00$ & $1.00\!\pm\!.00$ \\
two-shot (fixed) & $.66\!\pm\!.04$ & $.41\!\pm\!.04$ & $.74\!\pm\!.21$ & $.92\!\pm\!.02$ & $.92\!\pm\!.02$ \\
$+$adapter & $.77\!\pm\!.05$ & $.67\!\pm\!.03$ & $.68\!\pm\!.17$ & $1.00\!\pm\!.00$ & $1.00\!\pm\!.00$ \\
\midrule
positional copy & $1.00\!\pm\!.00$ & $.40\!\pm\!.00$ & $.44\!\pm\!.00$ & $1.00\!\pm\!.00$ & $1.00\!\pm\!.00$ \\
\bottomrule
\end{tabular}
\end{table}

Ordering support changes the interpretation of the per-case mean. At 0.5B, only 4 of
30 adapter generations match at least two reference steps, yielding only four comparable
pairs; its 0.73 ordering mean is therefore not evidence of ordered planning. Pair-weighted
ordering is 0.636 for adapter versus 0.667 for schema at 1.5B, and 0.742 versus 0.818 at
7B. We make no ordering-improvement claim.

Training-seed stability is measured separately with greedy decoding, so generation noise
does not masquerade as training variance. Across seeds 42--44, recall/precision is
$0.287\!\pm\!0.049$/$0.347\!\pm\!0.069$ at 0.5B,
$0.463\!\pm\!0.113$/$0.481\!\pm\!0.099$ at 1.5B, and
$0.806\!\pm\!0.042$/$0.727\!\pm\!0.098$ at 7B (sample standard deviation over three
adapters). Candidate validity is $0.981\!\pm\!0.032$, $0.972\!\pm\!0.024$, and
$1.000\!\pm\!0.000$. Three seeds are a stability diagnostic, not a population estimate.

\textbf{All-case case-sensitivity census.} We additionally run greedy leave-one-case-out (LOCO)
evaluation at 0.5B and 1.5B. Each of 24 folds trains a fresh seed-42 adapter on the other
23 cases; schema and two-shot baselines use the same fold, and exemplars never include the
test case. Table~\ref{tab:loco} therefore measures every authored case once while unseen in
training. At 0.5B, adapter and two-shot selection are close: recall/precision is
0.306/0.353 versus 0.321/0.335. Candidate validity differs more, 0.938 versus 0.717
(15/7/2 adapter win/tie/loss cases). At 1.5B, the adapter trades recall for selectivity:
precision is 0.583 versus 0.480 for two-shot (13/3/8), while recall is 0.586 versus
0.660 (5/7/12). Its ordering of 0.715 exceeds two-shot's 0.618 but not schema's 0.736.
A format-neutral extraction leaves 1.5B adapter/two-shot precision at 0.583/0.473 and
recall at 0.586/0.660, so that tradeoff is not a structured-parser artifact. This census
describes the 24 authored cases under case exclusion; it is not family-disjoint and supports
neither a population-transfer claim nor a general adapter advantage. The 7B scale comparison
is excluded from this conclusion because no 7B LOCO run was performed.

\begin{table}[t]
\centering
\caption{Greedy 24-fold case-level LOCO census at the two compact sizes. Each authored case
is evaluated by an adapter trained on the other 23 cases. R/P/O/G are recall, precision,
ordering, and candidate validity.}
\label{tab:loco}
\setlength{\tabcolsep}{4pt}
\scriptsize
\begin{tabular}{llcccc}
\toprule
\textbf{Size} & \textbf{Config} & \textbf{R} & \textbf{P} & \textbf{O} & \textbf{G} \\
\midrule
0.5B & schema   & .278 & .154 & .208 & .419 \\
     & two-shot & .321 & .335 & .667 & .717 \\
     & adapter  & .306 & .353 & .708 & .938 \\
\midrule
1.5B & schema   & .665 & .475 & .736 & .943 \\
     & two-shot & .660 & .480 & .618 & .958 \\
     & adapter  & .586 & .583 & .715 & 1.000 \\
\bottomrule
\end{tabular}
\end{table}

\textbf{Q3 (next-step continuation).} In the autoregressive condition, the first prompt
has no prior step and each later prompt contains only the model-generated prefix. This is a
continuation diagnostic, not an autonomous decomposition test: the reference length is
supplied as an explicit oracle horizon and stopping is not evaluated. Adapter technique accuracy is 0.08,
0.06, and 0.29 at 0.5B, 1.5B, and 7B, against 0.01, 0.02, and 0.04 for the base.
Absolute accuracy is poor, especially for the two smaller models. At 7B, teacher forcing
raises adapter accuracy to 0.37 (base 0.09), consistent with compounding prefix errors; this
condition was not run at the two smaller sizes and is not used for a cross-size claim.

\textbf{Transfer and execution diagnostic.} We reused the action--technique--check contract
on three public crackmes. An internally served GLM-5.2-NVFP4 arm produced 15/15 complete JSON responses
under clean reference grounding, but only 2/15 used the literal \code{check} key. A labeled
post-hoc key-normalizing diagnostic gives technique-sequence completeness 0.933 and
precision, ordering, supplied-set membership, and alias-normalized check-key presence 1.000;
exact-contract rate is $0.133\pm0.163$. An earlier direct crackme exercise motivated this
controlled rerun but is excluded from the reported evidence because its raw execution packet
is unavailable. The public evidence capsule contains all three prompts and references, 15
sanitized API requests and responses, seeds, sampling settings, and the deterministic
aggregator, so these values can be recomputed exactly. The serving endpoint exposed no
immutable weight revision or model fingerprint, so token-identical regeneration is not
claimed; the protocol can instead be rerun against a user-supplied compatible endpoint.
In a blinded exercise mapped to known CVE-2025-32433, our
isolated rerun retained full logs and reproduced a schema-valid vulnerable-positive/patched-
negative result for Erlang/OTP 27.3.2/27.3.3, including identical repeated vulnerable-image
builds. The capsule includes the digest-pinned internal-network lab, fixed-marker oracle,
environment record, logs, validation, and checksums. These results show that the three-field decomposition idea can be reused with
explicit normalization and verification; they do not show adapter transfer, novel CVE
discovery, or population generalization~\cite{satsecdataset2026}.

\subsection{Threats to validity}
\label{sec:threats}
The six fixed cases are development-facing and do not estimate a population; resampling
them only describes split sensitivity. LOCO is a census of the complete authored corpus rather
than an external sample, and related cases or technique families remain in other folds, so
it is neither family-disjoint nor evidence about a target population. Three training seeds expose variability but remain
too few to characterize an optimizer-induced distribution. Candidate sets guarantee inclusion of every reference technique and
exclude sibling-level distractors; neither retriever recall nor harder near misses are
measured. The references are collectively author-reviewed operationalizations rather than
independently certified semantic ground truth. Source resolution, field completeness,
candidate inclusion, live identifier/title checks, and leave-one-reference-case-out
sensitivity make their construction and result dependence inspectable, but cannot prove that
each authored action, order, or check is the only valid decomposition.
Exact identifier matching ignores action correctness, while the explicitly syntactic
check-field metric records presence rather than semantic validity or successful execution. Formatting
compliance is therefore a confound, not merely a presentation detail. Ordering is defined
only over matched steps and has little support at low recall. The next-step rollout uses
the reference length as a horizon oracle, so it does not measure completion detection.
Finally, no device profile supports a lightweight-hardware claim. Optional community
review, a genuinely untouched external test set, realistic retrieval, semantic action
evaluation, and on-device profiling remain future work.
The transfer diagnostic uses an unadapted model, clean grounding, labeled post-hoc key
normalization, three crackmes, and one known-CVE exercise; it is not an adapter baseline,
crackme-solving result, or external estimate. Its released responses make the reported
metrics auditable, but the unavailable immutable GLM-5.2-NVFP4 build prevents a claim of
token-identical model regeneration.

\subsection{Ethics and release risk}
The artifact contains dual-use descriptions of eavesdropping, replay, unauthenticated
commanding, spoofing, and debug-interface abuse. These behaviors are already public, but
structuring them may reduce the effort needed to organize that information. We release the
corpus because transparent prompts, failures, controls, and scoring support defensive
benchmarking; we do not release credentials, target coordinates, transmit parameters,
general-purpose exploit tooling, or live-system automation. The diagnostic capsule includes
only a known-CVE lab oracle hard-coded to two internal Docker service names and one fixed
marker-writing payload; it accepts neither an external host nor a caller-supplied command.
The model only selects from a supplied candidate
set and neither retrieves targets nor executes or verifies actions. All evaluation cases are
passive research, authorized-study abstractions, canonical standards patterns, or emulated
bench scenarios. Prompts restrict use to consented or emulated targets, and outputs require
human safety, legal, and technical review before any authorized test. The artifact includes
the full release-risk statement and failed outputs rather than presenting the adapter as an
operational capability.
The CVE rerun used an internal-only network and a fixed marker-only probe against pinned
versions; it accepted no caller-supplied command and contacted no external target.

\section{Related Work}
\label{sec:related}

\textbf{Decomposition and small-model specialization.} Least-to-most prompting first
decomposes a problem and then solves its subproblems sequentially~\cite{leasttomost2023}.
Model specialization has also transferred multi-step mathematical reasoning to models below
11B parameters~\cite{smallreasoning2023}. Our work does not claim a new general
decomposition algorithm: it defines and releases a security-specific supervised task in
which a compact model emits steps from an explicit technique set.

\textbf{Retrieval and candidate ranking.} Retrieval-augmented generation combines a
learned retriever with parametric generation and evaluates both on knowledge-intensive
tasks~\cite{rag2020}. LLMs have also been studied directly as listwise rerankers over
retrieved candidates~\cite{llmrerank2023}. Our controlled task is closer to reranking than
to end-to-end retrieval because every reference technique is inserted into $G(o)$. We call
the present sets oracle candidates and leave retriever evaluation to future work.

\textbf{Structured output and domain adaptation.} Grammar-constrained decoding can
guarantee structural validity without finetuning~\cite{gcd2023}, while domain-specific
security models use curated instruction tuning~\cite{cyberpal2024}. Our LoRA mechanism is
standard~\cite{lora2022}; the contribution is the dataset and measurement setup. The
formatting diagnostic shows that adaptation mainly improves structural compliance in some
conditions, so syntactic validity and semantic candidate selection must remain separate.

\section{Conclusion}
\label{sec:conclusion}

We define a candidate-set objective-decomposition task for development-time satellite
security and release its 24 authored decompositions, 83 derived next-step examples,
deterministic construction and audit tools,
raw generations, and case-level analysis. This is a controlled proof of concept: all six
fixed cases were inspected during development, every reference technique is present in the
oracle candidate set. The all-case LOCO census exposes a precision--recall tradeoff rather
than a general adapter advantage. Paired intervals, three-seed stability, sparse
ordering support, and a separate formatting diagnostic prevent the observed means from
supporting a general adapter-advantage claim. Autoregressive next-step accuracy is also too
low for stand-alone planning. The artifact instead provides an auditable starting point for
testing candidate selection separately from retrieval and verification. The small
reverse-engineering diagnostic supports contract reuse but reinforces the reliability boundary.
The authored references and their source boundaries are fully exposed for inspection;
community semantic review, a genuinely untouched external test set, realistic retrieval,
broader training replication, and on-device profiling remain necessary before reliability or
deployment claims would be warranted.

\section*{Generative AI Use}
Generative AI tools assisted with code scaffolding, dataset and manuscript review, and
language revision. The authors verified the dataset construction, semantic mappings,
experimental results, references, and final manuscript and take full responsibility for the
work.

\section*{Data and Code Availability}
The leakage-controlled dataset is available on Hugging Face at
\url{https://huggingface.co/datasets/paolocmo/satsec-decomposition} under
\href{https://doi.org/10.57967/hf/9586}{doi:10.57967/hf/9586}. The complete
reproducibility package, including dataset-construction and audit tools, training and
evaluation code, run manifests, raw predictions, the transfer-diagnostic evidence capsule,
model-artifact links, and analysis artifacts, is available at
\url{https://github.com/paoloo/satsec-decomposition/tree/v2.0.1}.

\printbibliography

\end{document}